\pdfoutput=1
\documentclass[%
reprint,
superscriptaddress,
 amsmath,amssymb,
 aps,
]{revtex4-2}

\usepackage[pdftex]{graphicx}% Include figure files
\usepackage{dcolumn}% Align table columns on decimal point
\usepackage{bm}% bold math
\usepackage{here}
\usepackage{float}
\usepackage{physics}
\begin{document}
\preprint{APS/123-QED}

\title{Nuclear norm regularized loop optimization for tensor network}% Force line breaks with \\

\author{Kenji Homma}
\affiliation{%
 Institute for Solid State Physics, The University of Tokyo, Kashiwa, Chiba 277-8581, Japan
}%

 \author{Tsuyoshi Okubo}
 \affiliation{%
 Department of Physics, The University of Tokyo, Tokyo 113-0033, Japan
 }%
 \author{Naoki Kawashima}%
 \affiliation{%
  Institute for Solid State Physics, The University of Tokyo, Kashiwa, Chiba 277-8581, Japan
 }%

\date{\today}% It is always \today, today,
    % but any date may be explicitly specified
\begin{abstract}
    We propose a loop optimization algorithm based on nuclear norm regularization for tensor network. The key ingredient of this scheme is to introduce a rank penalty term proposed in the context of data processing. Compared to standard variational periodic matrix product states method, this algorithm can circumvent the local minima related to short-ranged correlation in a simpler fashion. We demonstrate its performance when used as a part of the tensor network renormalization algorithms [S. Yang, Z.-C. Gu, and X.-G. Wen, Phys. Rev. Lett. 118, 110504 (2017)] for the critical 2D Ising model. The scale invariance of the renormalized tensors is attained with higher accuracy while the higher parts of the scaling dimension spectrum are obtained in a more stable fashion.

    \end{abstract}
\maketitle

%\tableofcontents
\section{Introduction}

Tensor network is a convenient representation of quantum and classical many-body problems in that effective truncation of degrees of freedom can be realized in a flexible way. Because of this capability, a number of promising approximation schemes have been proposed, as typified by Density Matrix Renormalization Group (DMRG) \cite{PhysRevLett.69.2863,PhysRevB.48.10345} and Corner Transfer Matrix Renormalization Group (CTMRG) \cite{doi:10.1143/JPSJ.65.891,doi:10.1143/JPSJ.66.3040}. Developments of progressive techniques such as Multi-scale Entanglement Renormalization Ansatz (MERA) \cite{PhysRevLett.99.220405} and projected
entangled pair states (PEPS) \cite{PhysRevB.90.064425,PhysRevLett.101.250602,PhysRevB.80.094403} would allow us to elucidate the complex many-body systems at finite bond dimension.

However, understanding the expressive power of tensor network for many-body systems is still limited to loop-free network. For bipartite tensor network ansatz, its gauge freedoms of loop-free tensor network can be fixed by introducing canonical gauges \cite{PhysRevLett.91.147902}. Its key idea was to find the center of orthogonality of tensor networks. This led to success of simulating infinite matrix product states (MPS) \cite{PhysRevLett.75.3537,PhysRevB.97.045145,PhysRevB.78.155117,PhysRevLett.91.147902} and generalized to any loop-free tensor networks. Yet, as Ref.\cite{PhysRevB.98.085155} discusses, there does not exist effective gauge-fixing method for looped tensor networks as bipartite canonical gauge is ill-defined for looped structure. Thus, developing efficient and computationally light method for training looped tensor network remains as a challenging task.

This issue can be traced back to short-range correlation problems in tensor renormalization group (TRG) \cite{PhysRevLett.99.120601}. Although TRG and its modified versions \cite{PhysRevB.86.045139,PhysRevLett.103.160601} have sufficient capability to compute partition functions of statistical models at moderate accuracies, Gu and Wen pointed out the TRG proposed by Levin-Nave (LN-TRG) does not produces correct RG flow because simple singular value decomposition (SVD) cannot remove the short-range correlations that lives in tensor space \cite{PhysRevB.80.155131}. Recently, tensor network renormalization (TNR) \cite{PhysRevLett.115.180405,PhysRevB.95.045117} and its variants \cite{PhysRevLett.118.110504,PhysRevLett.118.250602,PhysRevB.97.045111} were proposed to deal with such reduction of short-range correlations and produced more accurate results than TRG. For example, so-called Loop-TNR approach \cite{PhysRevLett.118.110504} has been formulated as ``loop optimization'', which is to decompose tensor network into loop-like configuration and variational method on minimizing the distance between two periodic MPS \cite{PhysRevB.81.081103,2008AdPhy..57..143V}.
Typically, it finds the minimum by solving linearization of cost function. However, such strategies will typically get stuck in local minima. Imposing symmetry restrictions to the search space, and/or trying better initial conditions are common practices to ease the problem. In the case of latter, one can start optimizing periodic MPS from the preconditioned points to accelerates the convergence. In general, a ``good'' preconditioned point highly depends on the underlying problem characteristics, which requires one to speculate the nature of local minima. We conjecture that the short-range correlation that lives in tensor space might be the origin of local minima for loop optimization.

In this paper, we propose alternative loop optimization algorithms, which we borrow from tensor ring decomposition techniques 
developed in the context of general data processing \cite{zhao2016tensor,yuan2018tensor}. Our strategy is to circumvent the 
local minima originating from short-range correlations by introducing a rank penalty term using nuclear norm regularization (NNR)
 \cite{candes2008exact,cai2008singular,Recht_2010}, which is closely related to least absolute shrinkage and selection operator
  (LASSO) regression but includes much broader applications \cite{jaggi2010simple}. It is known that the LASSO regression induces
   sparsity in solution, and similarly, the NNR induces sparsity of singular values, which corresponds to low-rank solution. 
   Our results suggest that adding these regularization term in tensor network's optimization is an effective approach for designing optimal entanglement structure. 
   We examine its performance using it as a part of LN-TRG approach and found that our strategy enables us to obtain the critical 
   fixed-point tensor with high accuracies but less human effort.

This paper is organized as follows. In Sect~\ref{methods}, we explain the target situation of NNR loop optimization and present its algorithmic details. In Sect~\ref{results}, we compare our proposal against conventional Loop-TNR. Thanks to NNR loop optimizations, short-ranged correlation are efficiently removed even at criticality and we find that our method further improves the scale-invariance of renormalization group, hence stabilizes higher parts of scaling dimension spectrum as a function of renormalization steps compared to Loop-TNR. A discussion and future applications of NNR loop optimization are presented in Sect~\ref{disc}.
\section{Loop optimization based on nuclear norm regularization}
\label{methods}
\begin{figure}
  %  \hspace*{-0.5cm}
    \includegraphics[scale=0.45]{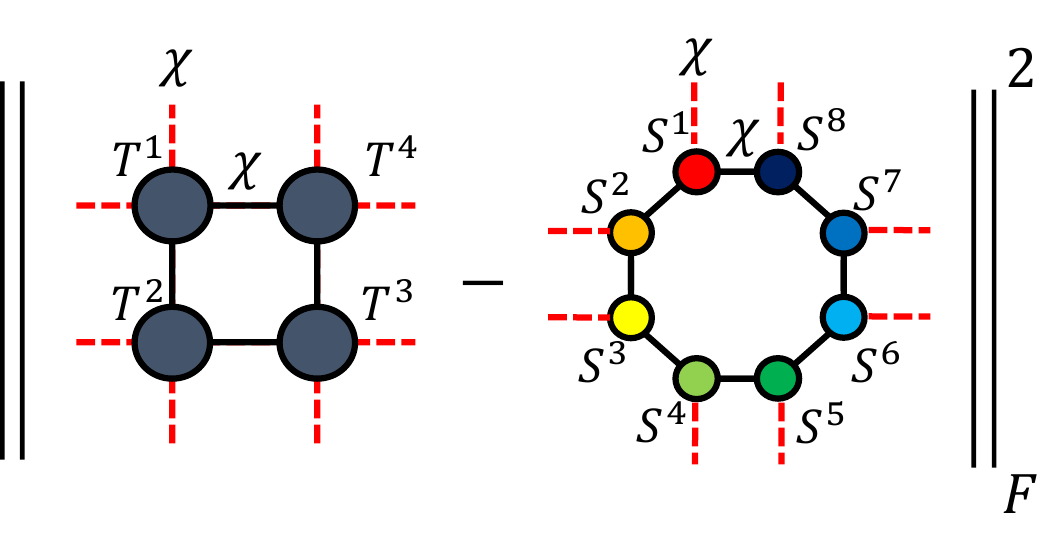}% Here is how to import EPS art
    \centering
   % \vspace{-1truemm}
    \caption{\label{fig:cost_function} The cost function of loop optimization. One can regard it as minimizing the distance between two different MPS. The red dotted line corresponds to physical indices and the black solid line are virtual indices of MPS with bond dimension $\chi$  }
\end{figure}

In the case presented below, we consider the following loop optimization problems in Fig.~\ref{fig:cost_function} for TNR application. Specifically,
\begin{equation}
 \label{eq.1}
||T^{1}\cdot T^{2} \cdot T^{3} \cdot T^{4}- S^{1} \cdot S^{2} \cdot S^{3} \cdot S^{4} \cdot S^{5} \cdot S^{6} \cdot S^{7} \cdot S^{8} ||^2_{F},
\end{equation}
where ``$\cdot$'' traces out the connecting indices in Fig.~\ref{fig:cost_function} and $||\cdots ||_{F}$ represents Frobenius norm. The square configurations of $T$ is now deformed into an octagon consisted of 3-index tensor $S$. In this paper, we shall call them as target and variational tensors, respectively. Typically, 3-index tensors are obtained by LN-TRG and treated independently. In addition, one can regard Eq.(\ref{eq.1}) as the minimizing the distance of two wavefunction with physical indices indicated by red dashed lines in Fig.~\ref{fig:cost_function} as $||\ket{\psi({\cal T})} -\ket{\psi({\cal S})}||^{2}_{F}$. Here, these calligraphic letters, for example $\mathcal{T}$, refer to the sets of tensors $\{T^{1},T^{2},\cdots \}$ to compose the periodic MPS $\ket{\psi(\mathcal{T})}$. Thus, the problem we concerned below is a general MPS optimization problem.

\subsection{Short-range correlations}
\begin{figure}
  %  \hspace*{-0.5cm}
    \centering
  %  \vspace{0truemm}
    \includegraphics[scale=0.25]{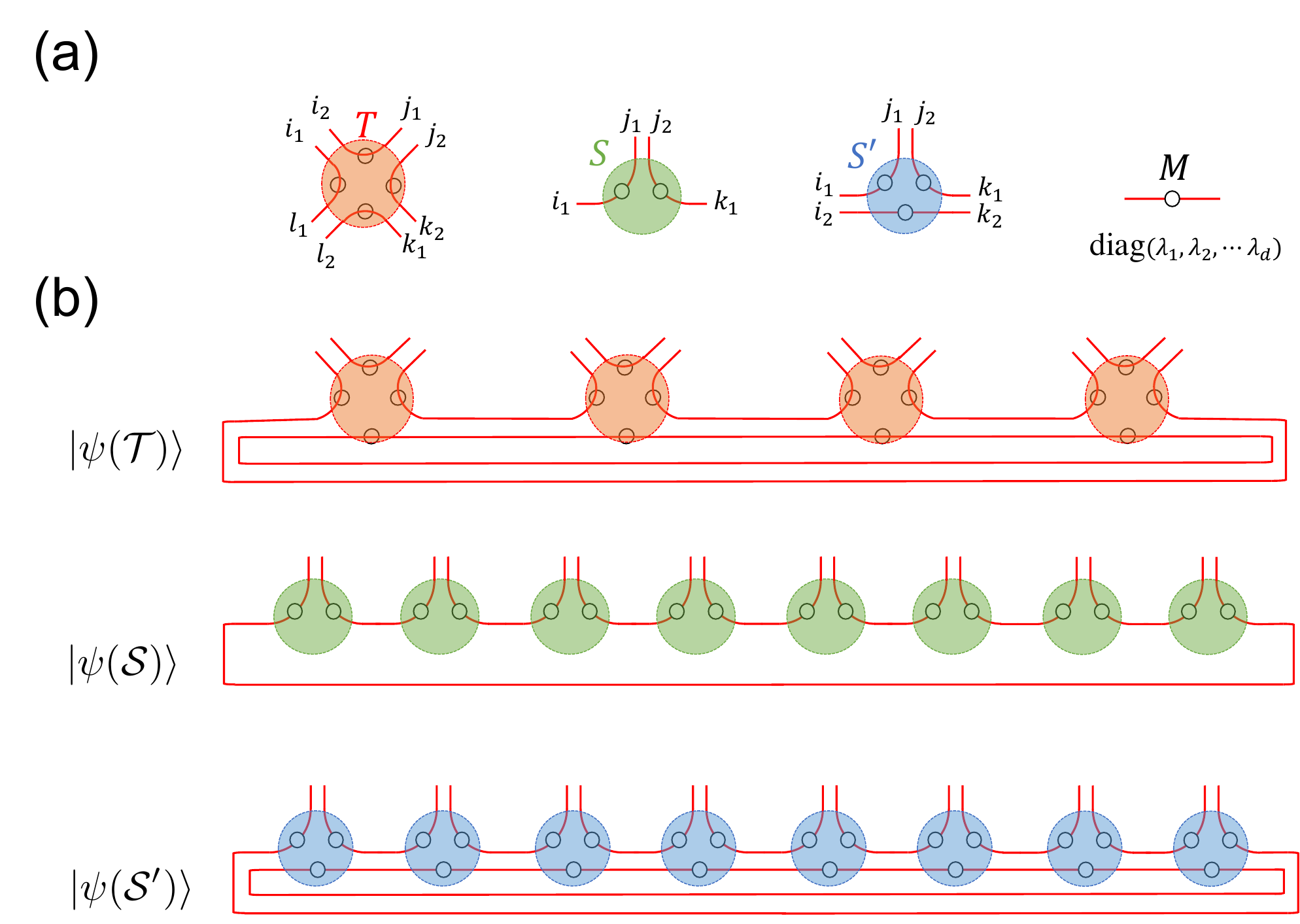}% Here is how to import EPS art
    \caption{\label{fig:cdl_mps} (a) Target CDL tensor $ T_{(i_{1}i_{2})(j_{1}j_{2})(k_{1}k_{2})(l_{1}l_{2})} $ and two examples of variational tensor $ S_{i_{1}(j_{1}j_{2})k_{1}}, S'_{(i_{1}i_{2})(j_{1}j_{2})(k_{1}k_{2})}$. (b) periodic MPS wavefunctions composed of $T,S$ and $S'$ under translational invariance.}% Note that these wavefunctions $\ket{\psi(\mathcal{T})}.$}%, $\ket{\psi(\mathcal{S})}$ and $\ket{\psi(\mathcal{S'})}$ are equivalent to each other.}
\end{figure}

To illustrate the algorithm, we consider the following special case where the target tensor $T$ in the Fig.~\ref{fig:cdl_mps}(a) has the form as,
\begin{eqnarray}
    \label{eq.2}
    T_{(i_{1}i_{2})(j_{1}j_{2})(k_{1}k_{2})(l_{1}l_{2})} = M_{i_{2}j_{1}}M_{j_{2}k_{2}}M_{k_{1}l_{2}}M_{l_{1}i_{1}},
\end{eqnarray} 
where $M= \operatorname{diag}(\lambda_{1},\lambda_{2}, \cdots , \lambda_{d})$ and $\lambda_{1}>\lambda_{2}> \cdots >\lambda_{d} $. Each subscripts of $T$, such as $(i_1,i_2)$, represents a single natural number, e.g., $i = (i_1,i_2) = i_1 + \eta (i_2-1)$ with $i_1,i_2 = 1,2,\cdots,\eta$. The integer $\eta$ is related to the bond dimension by $\chi = \eta^2$. The tensor $T$ is called corner double line (CDL) tensor, which is a toy model for short-ranged correlation. 
In addition, let us define a two different variational tensors $S$ and $S'$ for discussion.
\begin{eqnarray}
\label{eq.31}
 S_{i_{1}(j_{1}j_{2})k_{1}}&=& M_{i_{1}j_{1}}M_{j_{2}k_{1}},\\
 S'_{(i_{1}i_{2})(j_{1}j_{2})(k_{1}k_{2})} &=& M_{i_{1}j_{1}}M_{j_{2}k_{1}}M_{k_{2}i_{2}},
\end{eqnarray}
which follows index convention we define above. One can construct periodic MPS $\ket{\psi(\mathcal{T})}, \ket{\psi(\mathcal{S})}$, and  $\ket{\psi(\mathcal{S'})}$ from copies of $T$, $S$ and $S'$ under the assumption of translational invariance, as shown in Fig.~\ref{fig:cdl_mps}(b).  The first things to note is that the two wavefunctions $\ket{\psi(\mathcal{S})}$ and $\ket{\psi(\mathcal{S'})}$ are identical to each other up to some normalization factor, as the inner correlation loop formed by $M_{i_{2}k_{2}}$ can be contracted to an overall multiplicative factor. In this sense, the wavefunction  $\ket{\psi(\mathcal{S'})}$ has redundant correlation compared to  $\ket{\psi(\mathcal{S})}$.  Furthermore, the objective function in Eq.(\ref{eq.1}) gives $||\ket{\psi({\cal T})} -\ket{\psi({\cal S})}||^2_F = ||\ket{\psi({\cal T})} -\ket{\psi({\cal S'})}||^2_F = 0$, which implies that one cannot distinguish them from Eq.(\ref{eq.1}). The key issues in loop optimizing this periodic MPS can be addressed in indistinguishably of relevant correlation and redundant correlation \cite{PhysRevB.98.085155}, and we speculate it is the origin of local minima for loop optimization.
In Loop-TNR approach, there are several methods to implicitly avoid such local minima related to redundant correlation, such as so-called ``entanglement filtering'' and imposing the rotational lattice symmetry during the loop optimization. In this work, we propose alternative approach to penalize such redundant correlation via low-rank regularization.
\subsection{NNR loop optimization}
\begin{figure}
    \centering
    \includegraphics[scale=0.42]{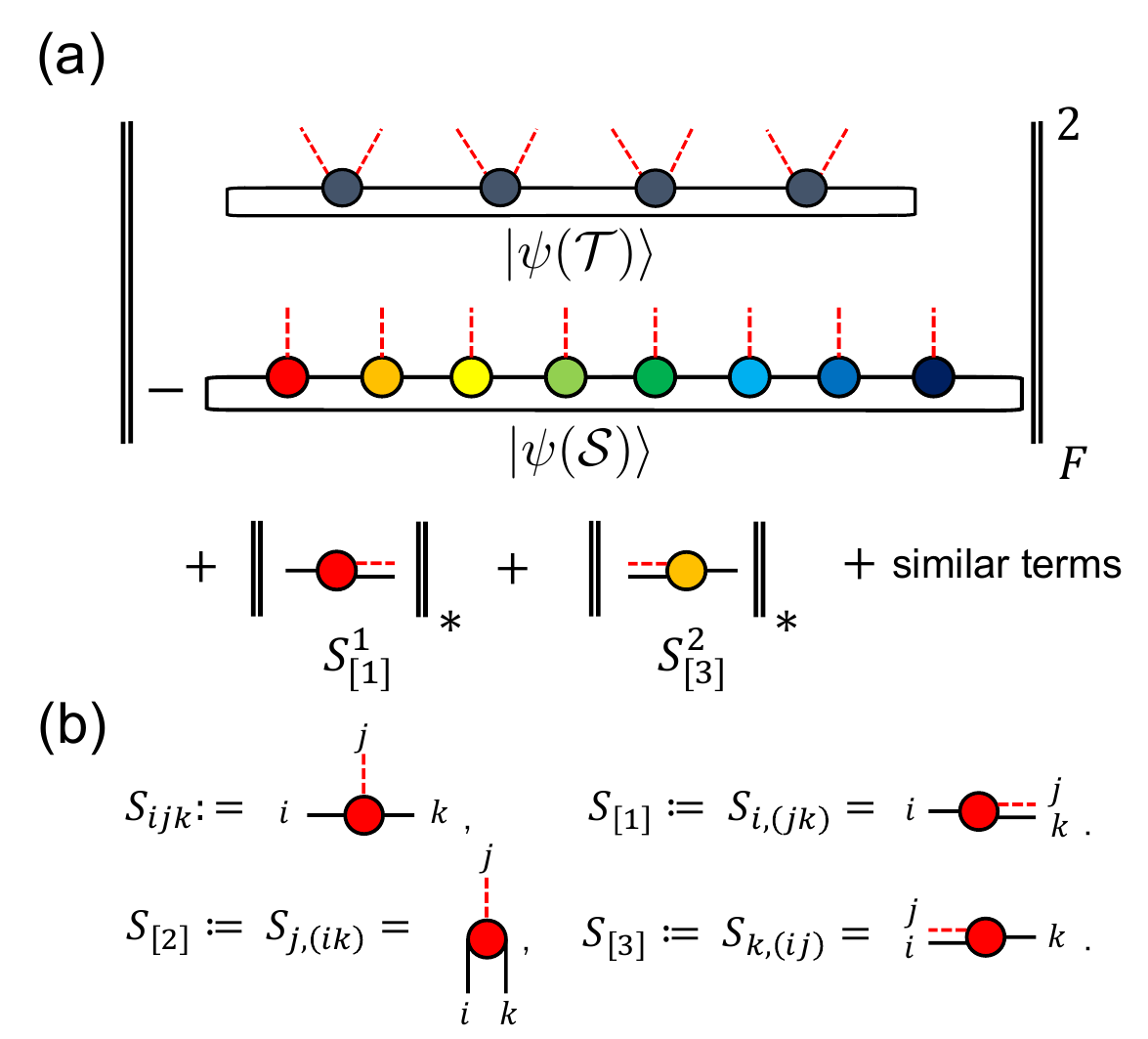}% Here is how to import EPS art
  %  \vspace{-2truemm}
    \caption{\label{fig:NNR_cost1}(a) The cost function of NNR loop optimization. (b) mode-$\alpha$ matricization. The lower indices $S_{[\alpha]}$ indicate which indices to be grouped.}
\end{figure} 
In the following, we consider penalizing the high-rank contributions of variational tensors, then gradually relax the penalty term to the original cost function Eq.(\ref{eq.1}).
However, the rank of a tensor is not convenient for the penalty term, because it is an integer and also because it is a rather complicated function of the tensor elements.
To overcome this problem, one of commonly used measures in rank regularization approach is the {\it nuclear norm}, which is defined as the sum of singular values. 
The NNR is a concept that has a broad area of possible applications, e.g., the basis pursuit de-noising problem in compressed sensing literature, 
and the LASSO regression popular in machine learning \cite{jaggi2010simple}. The fact that the singular values enter as their first power 
makes the whole algorithm stabler in convergence to lower rank solution, as we see in the LASSO regression.

%Apart from the NNR, another valid regularization approach to induce low rankness is entropy regularization, which penalizes the entropy of singular values. While both regularization techniques could serve as entanglement filtering, we believe that NNR is more efficient, at least for the TNR method, since it forces the small singular values to zero. The concrete comparisons between NNR and entropy regularization in the context of TNR are discussed in Ref.\cite{SM_NNR}.
While we penalize the high-rank representation through the NNR, there may be several other ways for the same purpose. Among them, we tried local entropy, 
or mutual information, as a possible alternative to the nuclear norm as we discuss in detail in the supplemental material \cite{SM_NNR}. It turned out that estimation of 
the central charge and scaling dimensions are noisier and less stable compared to the nuclear norm, while the free energy precision is comparable or even better in some cases. 
Though we have not identified the reason of this observations, we suspect that the singular nature of the nuclear norm may play an important role in improving the estimation of 
critical properties.

The method discussed in this section is a straight-forward application of the NNR loop optimization for tensor ring decomposition
proposed in Refs.\cite{zhao2016tensor,yuan2018tensor}, yet differs in the following two aspects: One is to dynamically adjust the amplitude
of the penalty term so that it would not spoil the high precision. This treatment is relevant because when the redundant correlation 
loop is only approximately redundant, i.e., when the redundant correlation is still weakly coupled to the other relevant degrees of 
freedom, which is usually the case in early iterations of the RG procedure, it is harmful to enforce the low-rank representation by 
the penalty term. The other aspect is that we incorporate constraints on the indices to penalize redundant correlations on the diagonal legs 
in Fig.~\ref{fig:cost_function}, which is to enhance consistency with Ref. \cite{PhysRevB.80.155131}.

Below, we detail how we define the penalty term using low-rank regularization for periodic MPS. The objective function under nuclear 
norm regularization in the Fig.~\ref{fig:NNR_cost1}(a) is given by
\begin{eqnarray}
\label{eq.3}
\min_{\mathcal{S}} \frac{1}{2}||\ket{\psi(\mathcal{T})}- \ket{\psi(\mathcal{S})} ||^{2}_{F} + \sum_{n=1}^{N} \mu || S_{[\alpha(n)]}^{n} ||_{*}.
\end{eqnarray}
Here, $\mu$ is a non-negative parameter and $||\cdots||_{\ast}$ denotes the nuclear norm, i.e., $||S^n_{[\alpha(n)]}||_{\ast} \equiv \sum_i \sigma^n_{\alpha(n),i}$ where
 $\sigma^n_{\alpha(n),i}$ are the $i$ th singular value of the $\alpha$-th mode ``matricized'' tensor $S^n_{[\alpha(n)]}$ at the $n$-th site. For example, $S_{[1]}$ is the matrix such that $S_{[1]} := S_{i,(jk)}$ where the doublet 
 $(jk)$ represents, as before, a single integer that uniquely specifies both $j$ and $k$ (see also Fig.~\ref{fig:NNR_cost1}(b)). $\alpha(n)$ is the 
 index corresponding to the diagonal legs in Fig.~\ref{fig:cost_function}, e.g. for $S^1$, $\alpha(n)$ specifies the leg connecting $S^1$ to $S^2$, and for $S^8$ it specifies the 
 leg connecting to $S^7$. To achieve this, we define the index function $\alpha(n)$ as $\alpha(n) = 1$ for $n$ being odd, and $\alpha(n) = 3$ for $n$ being even.
 Note that we assume, for later convenience, 
each target tensor $ T $ has been normalized, such that $\bra{\psi(\mathcal{T})}\ket{\psi(\mathcal{T})} = 1$. The initial value of variational tensor $S$ is computed from LN-TRG. 
The second term penalizes the nuclear norm of matrices $S_{[\alpha(n)]}$, which is supposed to restrict the high-rank solutions of $\ket{\psi(\mathcal{S})}$ and thus favors compact 
tensor network representations during optimization. 

In what follows, for each 3-index tensor $S^{n}$, we introduce another sets of 3-index tensors $M^{n}$ as auxiliary variables to handle the first and second terms in Eq.(\ref{eq.3}) independently, which is given by
\begin{align}
\begin{split}
\label{eq.4}
\min_{\mathcal{S},\mathcal{M}} &\ \frac{1}{2}||\ket{\psi(\mathcal{T})}- \ket{\psi(\mathcal{S})}||^{2}_{F} + \sum_{n=1}^{N} \mu||M_{[\alpha(n)]}^{n}||_{*} \\
&\mathrm{s.t.}\ \ M^{n}_{[\alpha(n)]} = S^{n}_{[\alpha(n)]}.
\end{split}
\end{align}
As is discussed below, while $M^{n}_{[\alpha(n)]}=S^{n}_{[\alpha(n)]}$ will be achieved after sufficient iteration, we let them differ from 
each other during the optimizations. Incorporating the condition in Eq.(\ref{eq.4}) in the style of alternating direction 
method of multipliers (ADMM), the augmented Lagrangian can be written by introducing the Lagrange multipliers $Y^{n}$ to 
impose the condition $M^{n}_{[\alpha(n)]}=S^{n}_{[\alpha(n)]}$, as
\begin{eqnarray}
 \label{eq.5}
\mathcal{L}(\mathcal{S},\mathcal{M},\mathcal{Y}) &=& \frac{1}{2}||\ket{\psi(\mathcal{T})}-\ket{\psi(\mathcal{S})}||^{2}_{F}
 + \sum_{n=1}^{N}\mu ||M_{[\alpha(n)]}^{n}||_{*} \nonumber \\
 &+& \mathrm{tr}(Y_{[\alpha(n)]}^{n}(M_{[\alpha(n)]}^{n} -S^{n}_{[\alpha(n)]})) \nonumber \\
 &+& \frac{\xi}{2}||M_{[\alpha(n)]}^{n}-S^{n}_{[\alpha(n)]}||^{2}_{F},
\end{eqnarray}
where $0< \xi < 1$ is a penalty parameter. The initial conditions of $M^{n},Y^{n}$ are zero-elements tensors. Update equations of given augmented Lagrangian can be obtained by the extremum of Eq.(\ref{eq.5}) with respect to $S^{n}$ and $M^{n}$ for $n = 1,2,\cdots,N$. In what follows, we explain each update equations of $ S^{n}, M^{n}, Y^{n}$ for single iteration.
\paragraph{Solving Scheme for $ S^{n}$}
\begin{widetext}
One can simplify the augmented Lagrangian function of ${S}^{n}$ as
\begin{eqnarray}
 \label{eq.6}
\mathcal{L}(S^{n}) &=& \frac{1}{2}|| \ket{\psi(\mathcal{T})}- \ket{\psi(\mathcal{S})} ||^{2}_{F}%\nonumber \\
 + \sum_{n=1}^{N} \mathrm{tr}(Y_{[\alpha(n)]}^{n}(M_{[\alpha(n)]}^{n} -S^{n}_{[\alpha(n)]})%\nonumber \\
 + \frac{\xi}{2} ||M_{[\alpha(n)]}^{n} -S^{n}_{[\alpha(n)]}||^{2}_{F}).
\end{eqnarray}
The first quadratic term in th Eq.(\ref{eq.6}) can be expanded in terms of $S^{n}$ as

\begin{eqnarray}
|| \ket{\psi(\mathcal{T})}- \ket{\psi(\mathcal{S})} ||^{2}_{F} &=&   \bra{\psi(\mathcal{T})}\ket{\psi(\mathcal{T})} + \bra{\psi(\mathcal{S})}\ket{\psi(\mathcal{S})}  
-\bra{\psi(\mathcal{S})}\ket{\psi(\mathcal{T})}-\bra{\psi(\mathcal{T})}\ket{\psi(\mathcal{S})}  \nonumber \\
&=&  1 + \sum_{ijklm}(S^{n}_{ijk})^{\dagger}N^{n}_{iklm}S^{n}_{ljm}
-S^{n}_{ljm}(W^{n}_{ljm})^{\dagger} -(S^{n}_{ijk})^{\dagger}W^{n}_{ijk},
\end{eqnarray}
or diagrammatically
\begin{figure}[H]
    \centering
    \vspace*{-0.2cm}
    \includegraphics[scale=0.4]{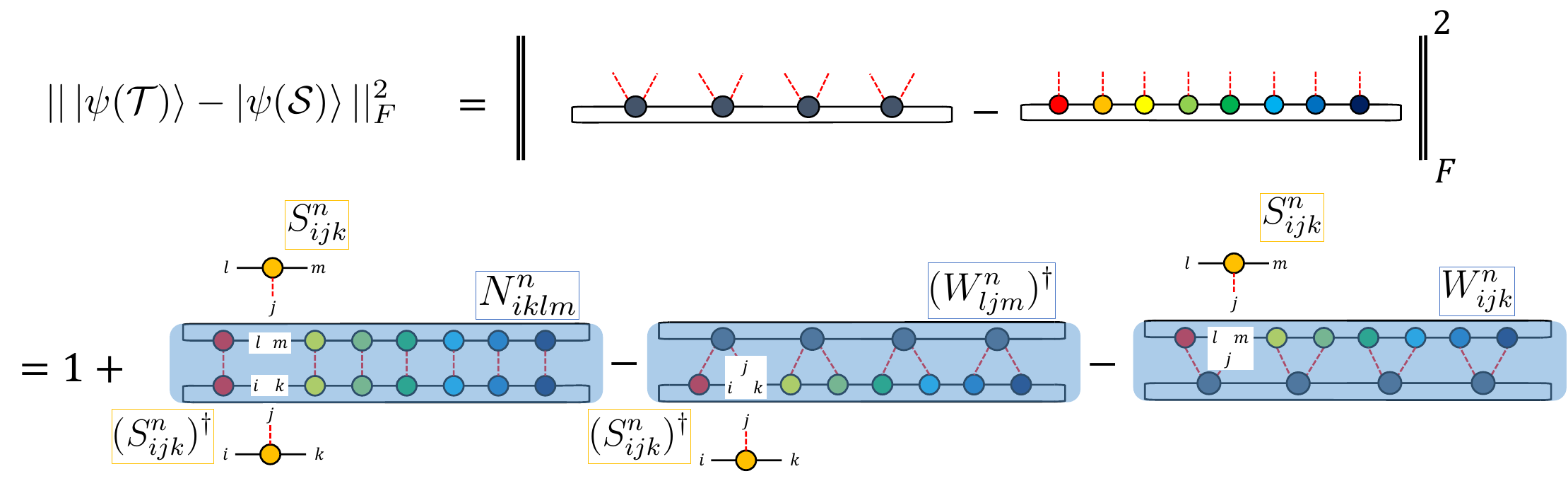}% Here is how to import EPS art
\end{figure} 
Suppose all tensors but $S^{n}$ are fixed, the minimum of Eq.(\ref{eq.6}) can be obtained by solving the linear equation 
\begin{eqnarray}
\label{eq.7}
({N^{n}_{(ik)(lm)} +\xi I})(S^{n}_{[2]})^{\dagger} &=& W^{n}_{[2]}+Y^{n}_{[\alpha(n)]}+\xi M^{n}_{[\alpha(n)]},
\end{eqnarray} 
where $ I \in \mathbb{R}^{\chi^{2} \times \chi^{2}}$ denotes identity matrix. One can update as $S^{n} \leftarrow tensorize(S^{n}_{[2]})$ for $n =1,2 \cdots N$. Here, $tensorize$ denotes tensorization of given input matrices into $3$-index tensor.
We used the least squares method to solve the linear equations here. The cutoff in this case is the machine epsilon.
\end{widetext}

\paragraph{Solving Scheme for $ M^{n}$}
For optimization of auxiliary variable $M^{n}$, its augmented Lagrangian can be written by
\begin{eqnarray}
 \label{eq.8}
\mathcal{L}(M^{n}) &=& \sum_{n=1}^{N} \xi ||M_{[\alpha(n)]}^{n} -S^{n}_{[\alpha(n)]} + \frac{1}{\xi}Y_{[\alpha(n)]}^{n}||^{2}_{F} \nonumber \\
&+& \mu || M_{[\alpha(n)]}^{n} ||_{*}.
\end{eqnarray}
The optimization problem for the augmented Lagrangian in Eq.(\ref{eq.8}) can be solved using singular value shrinkage operator (SVSO) \cite{cai2008singular} as shown below. For the sake of simplicity, we first consider the following problem under nuclear norm regularization.
\begin{eqnarray}
\min_{X} \tau||X||_{*}+ \frac{1}{2}||X-Y||^{2}_{F} ,\nonumber
\end{eqnarray}where $X,Y\in \mathbb{R}^{2} $ are unknown and given matrices respectively.
The SVSO $\mathcal{D_{\tau}}(Y)$ finds solution $X$ of above equations as
\begin{eqnarray}
\mathcal{D_{\tau}}(Y) := U \lfloor (s-\tau) I \rfloor V \nonumber,
\end{eqnarray}
where $\lfloor X_{ij} \rfloor \equiv \max(X_{ij},0)$ is the element-wise floor function of the matrix $X$. $U, s, V$ are obtained from the SVD of $Y$.
Thus, the updating equations for Eq.(\ref{eq.8}) is given by
\begin{eqnarray}
 \label{eq.9}
M^{n} = tensorize(\mathcal{D_{\mu/\xi}}(S^{n}_{[\alpha(n)]}))
\end{eqnarray}
for $n =1,2,\cdots N$.

\paragraph{Updating scheme of $ Y^{n} $}
Lastly, the Lagrangian multiplier $ Y^{n}  $ is updated as
\begin{eqnarray}
 \label{eq.10}
Y^{n} \leftarrow Y^{n} +  \xi(M^{n} - S^{n}).
\end{eqnarray}
The aforementioned three steps are applied sequentially to each of the $n$-th site, after which the optimization moves to the next site. Once this sweep has completed a full cycle, it constitutes a single iteration. 
Moreover, the penalty parameter $\xi$ is updated as $\xi \leftarrow \rho\xi$ at every iterations by $0 <\rho < 1$, such that the regularization term in Eq.(\ref{eq.5}) is gradually relaxed to the original Eq.(\ref{eq.1}). 
We found that by setting $\mu := \xi^2$ for the next iteration it works appropriately for all examples in Sect~\ref{results}. 

In summary, the NNR loop optimization has several parameters for users to tune. The larger $\xi$, our objective functions favor low-rank solutions and $\rho$ controls the penalty schedule. For the examples shown below, we chose the initial value of the parameter $\xi$, as denoted as $\xi^{\text{init}}$, and the parameter $\rho$ within ranges $10^{-7}< \xi^{\text{init}} <10^{-4}$ and  $0.8 <\rho <1$, respectively.
%\end{figure}
\section{Benchmark results of NNR loop optimization}
\label{results}
\begin{figure}
    \centering
    \includegraphics[scale=0.35]{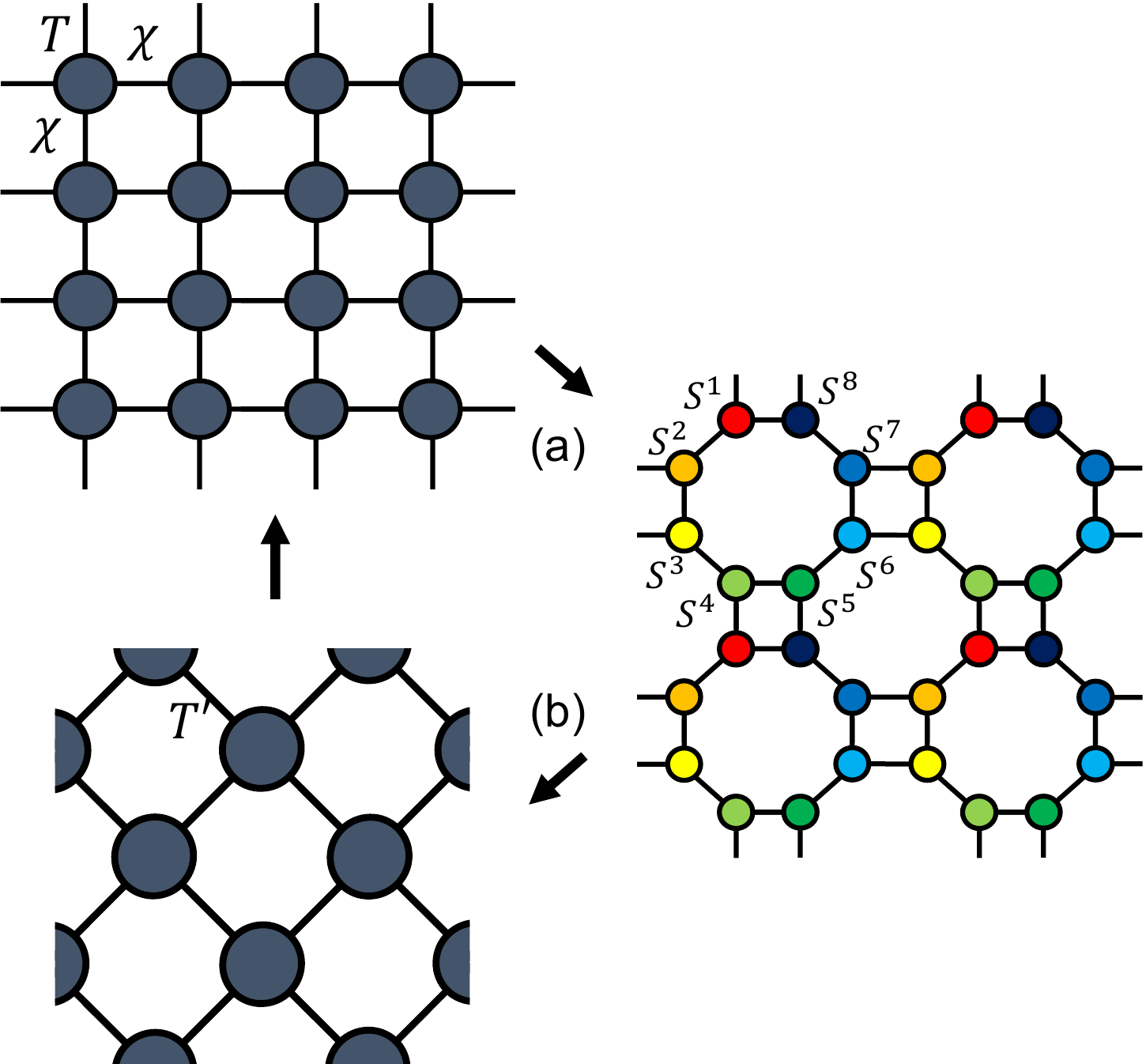}% Here is how to import EPS art
    \caption{\label{fig:nnr} Single iteration of NNR-TNR for a square lattice. (a) Target tensors $T$ are first decomposed into 3-index tensors using LN-TRG, then NNR loop optimization is applied to the 8 tensors on loop $S^{1}\cdot S^{2}\cdot S^{3} \cdot S^{4} \cdot S^{5} \cdot S^{6} \cdot S^{7}\cdot S^{8}$, as introduced in the previous section. (b) The updated 4 tensors are contracted into single tensor in the same fashion as LN-TRG. New target tensors are recovered from rotating the square lattice by $45^\circ$}
\end{figure}

In this section, we show an example application of NNR loop optimization for TRG. We call the combination of NNR loop optimization and TRG, {\it NNR-TNR}.
 One coarse-graining iteration for square lattice is shown in Fig.~\ref{fig:nnr}. At the first step [Fig.~\ref{fig:nnr}(a)], we compute the initial value of 
 the variational tensor $S$ from LN-TRG and perform NNR loop optimization as described in the previous section. 
Note that such a choice for the initial value of $S$ is quite useful for stable convergence, though LN-TRG does not remove redundant correlation.
The coarse-graining steps in Fig.~\ref{fig:nnr}(b) are the same as LN-TRG, as sets of 4 updated tensors $S$ are contracted to yield new target tensor $T'$. 
As will be discussed later in this section, NNR loop optimization filters out the redundant correlation during optimization, and thus it not only optimizes 
the approximation given by Eq.(\ref{eq.1}) but also has an effect similar to the entanglement filtering of Loop-TNR. In this sense, NNR-TNR achieves high-accuracy 
results with less human efforts, as its minimal programming implementation takes much fewer lines of codes than conventional Loop-TNR (See App.~\ref{app1}). 
The overall computational complexity scales as $O(\chi^{6})$ if one follows the contraction order given in Ref.\cite{PhysRevLett.118.110504}.

We benchmark NNR-TNR by several physical observable for the critical 2D Ising model on the square lattice. Note that our formulism does not impose any symmetries, such as $Z_{2} $ symmetry, 
rotational or reflectional symmetries of 8 octagon tensors for all the results below. Therefore, we compare the results of NNR-TNR with those of Loop-TNR without imposing the rotational 
symmetries of lattices \footnote[2]{For our implementation of Loop-TNR, we have performed variational periodic MPS method and entanglement filtering, as proposed in the original literature. 
In addition, we have chosen the initial value of variational periodic MPS from the method proposed in Ref.\cite{wang2011cluster} to accelerate the convergence. Our implementation is 
supposed to faithfully reproduce the results in Ref.\cite{PhysRevLett.118.110504}}.

In tuning the parameters of NNR-TNR methods for analyzing critical systems, we observed the following trends. When the penalty parameter is set too low, contributions from short-range 
correlations persist, leading to instability in the RG flow and preventing improvements in accuracy even with increasing bond dimension. Conversely, if the penalty parameter is set too high, 
not only the short-range correlations but also the relevant information will be truncated. Consequently, the RG flow rapidly converges to a trivial fixed point. For effective 
parameter tuning in NNR-TNR, it is recommended to adjust the penalty parameters to maintain scale-invariant fixed-point tensor up to the largest possible system sizes. 
The end of RG flow can be checked from the singular value spectrum, as described in the following paragraph. Note that the parameters should ideally be adjusted according to bond dimension.

%\begin{figure}
%    \centering
%    \includegraphics[scale=0.5]{free_energy.pdf}% Here is how to import EPS art
%    \caption{\label{fig:free_ene} Relative error of free energy density of critical 2D classical Ising model as a function of bond dimension using Loop-TNR and NNR-TNR with 
%    $2^{30}\times 2^{30}$ spins. The relative error is computed against the exact free energy density at the thermodynamic limit. The blue solid line corresponds to NNR-TNR result with the parameters 
%    $\xi^{\text{init}}=1\cdot 10^{-5}$ and $\rho=0.84$, chosen to be fixed over the full range of bond dimensions for simplicity. Colorful dots are the results of NNR-TNR using optimal parameters 
 %   tuned for each bond dimension.
%    }
%\end{figure}

\begin{figure}
    \includegraphics[scale=0.5]{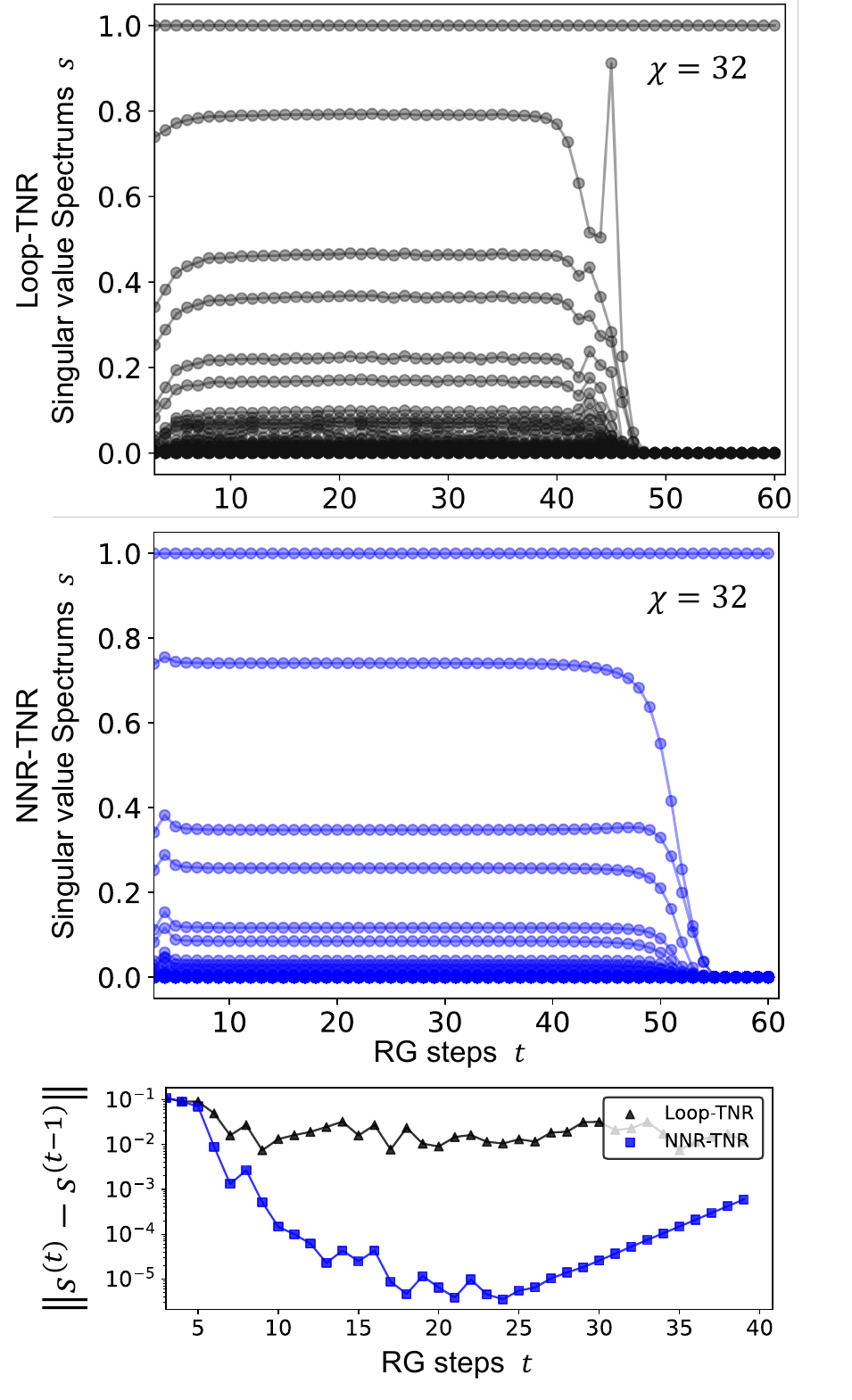}% Here is how to import EPS art
    \caption{\label{fig:spectrum}(Top and Middle Panel) Singular value spectrums $s^{(t)}$ of the coarse-grained tensors $T^{(t)}$, illustrated for Loop-TNR (Top) and NNR-TNR (Middle), at the exact critical temperature of 2D Ising model for different RG steps $t$, which is equivalent to linear system size $2^{t}$. At each RG steps, the largest 100 singular values of tensors $T$ are displayed. The largest number of $s^{(t)}$ is normalized to 1 for all $t$. (Bottom) The distance between singular value spectrums $s^{(t)}$ and $s^{(t-1)}$ for both methods. This distance corresponds to the order of the scale-invariance at each RG steps. In this figure, the parameters of NNR-TNR have been chosen as  $\xi^{\text{init}} = 4\cdot 10^{-6}$ and $\rho = 0.85$.
    }
\end{figure}

%In Fig.~\ref{fig:free_ene}, we show the relative error of free energy density of critical 2D Ising model at the thermodynamic limit with various bond dimension $\chi$, compared with Loop-TNR and NNR-TNR. For $\chi \geq 8$, it seems NNR-TNR generates slightly more accurate results than Loop-TNR. At $\chi = 24$, NNR-TNR obtains  $\delta f \approx 1\cdot 10^{-10}$, which is comparable to the best previous studies \cite{PhysRevLett.118.110504,PhysRevLett.115.180405,PhysRevB.98.085155}. Note that we have fixed the parameters of NNR-TNR  as $\xi^{\text{init}} =4\cdot 10^{-6}$ and $\rho = 0.85$ for the blue solid line in Fig.~\ref{fig:free_ene}. Further hyperparameter tuning may generate more accurate results with smaller bond dimension.

In Fig.~\ref{fig:spectrum}, we present singular value spectrums of coarse-grained tensors, which supposedly monitors the RG flow at the unstable critical fixed-point. 
For Loop-TNR, it obtains approximately identical singular value spectrums as a function of the RG steps starting from the 3rd iteration, roughly up to the 40th. 
In contrast, NNR-TNR maintains the almost identical unstable critical fixed-point tensors more than 40th RG steps. In addition, the RG flow ends in the trivial fixed-point 
tensor for both results, which implies that there is no footprint of short-range correlation for NNR-TNR, as well as Loop-TNR. Thus, it seems that NNR-TNR successfully truncates 
the short-range correlation during loop optimization. The bottom panels of Fig.~\ref{fig:spectrum} shows the approximations to scale-invariance measured 
by $\delta = ||s^{(t)}-s^{(t-1)}||$, where $s^{t}$ is the singular values of the normalized tensor $T^{(t)} = Us^{(t)}V^{\dagger}$ at the $t$-th RG step. 
It seems that NNR-TNR improves Loop-TNR in terms of fixing the scale-invariant tensor at criticality.

\begin{figure}
    %\hspace*{-0.5cm}
    \includegraphics[scale=0.445 ]{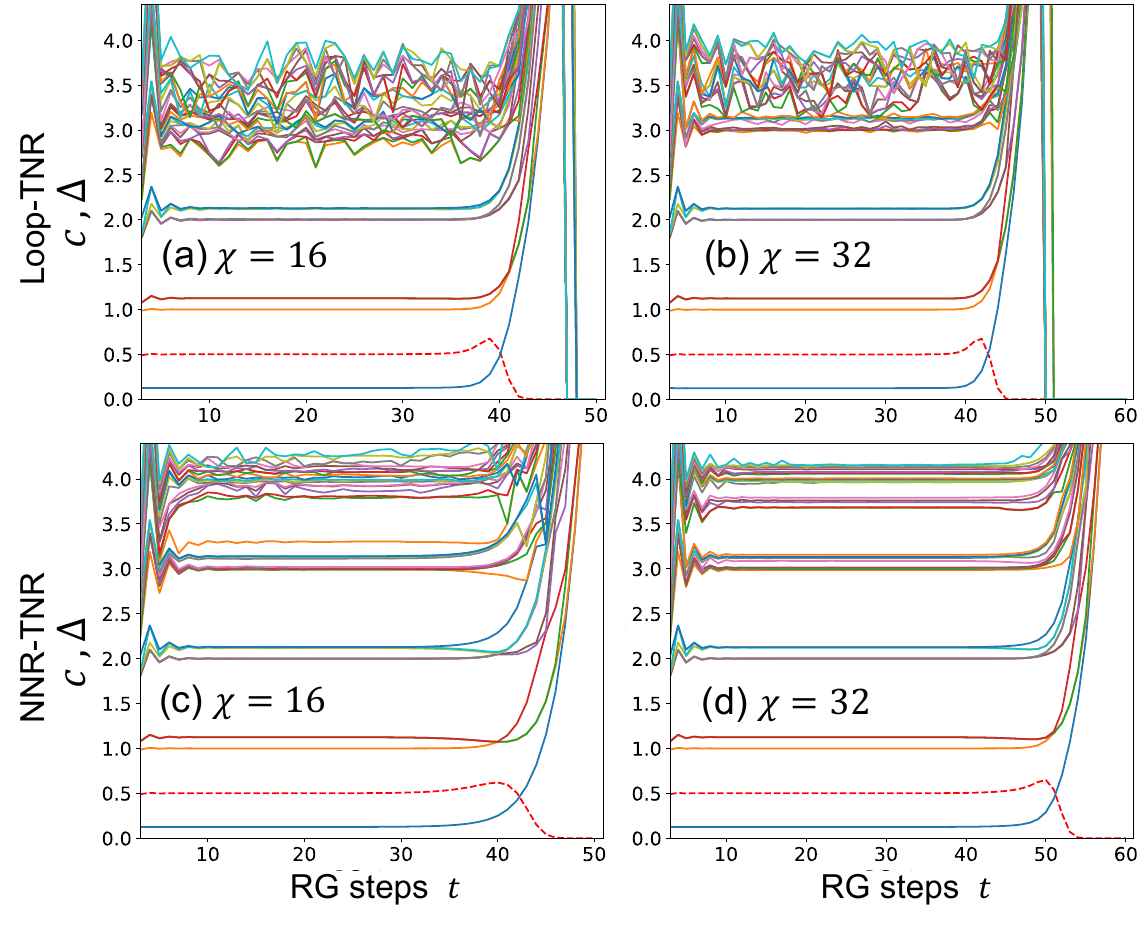}% Here is how to import EPS art
    \caption{\label{fig:cft}Comparison of central charges (red dotted line) and scaling dimensions (solid lines) for Loop-TNR and NNR-TNR at 
    different RG steps. Results are computed by Gu and Wen’s methods \cite{PhysRevB.80.155131} using $L=2$ transfer matrix in 
    Ref.\cite{PhysRevLett.118.110504}. We chose the parameters of NNR-TNR for (c) and (d) have been chosen as $\xi^{\text{init}}= 1\cdot 10^{-5}, \rho = 0.8$ 
    and  $\xi^{\text{init}}= 4\cdot 10^{-6}, \rho = 0.85$, respectively.
     }
\end{figure}
\begin{figure}
    %    \vspace*{-0.5cm}
        \includegraphics[scale=0.45]{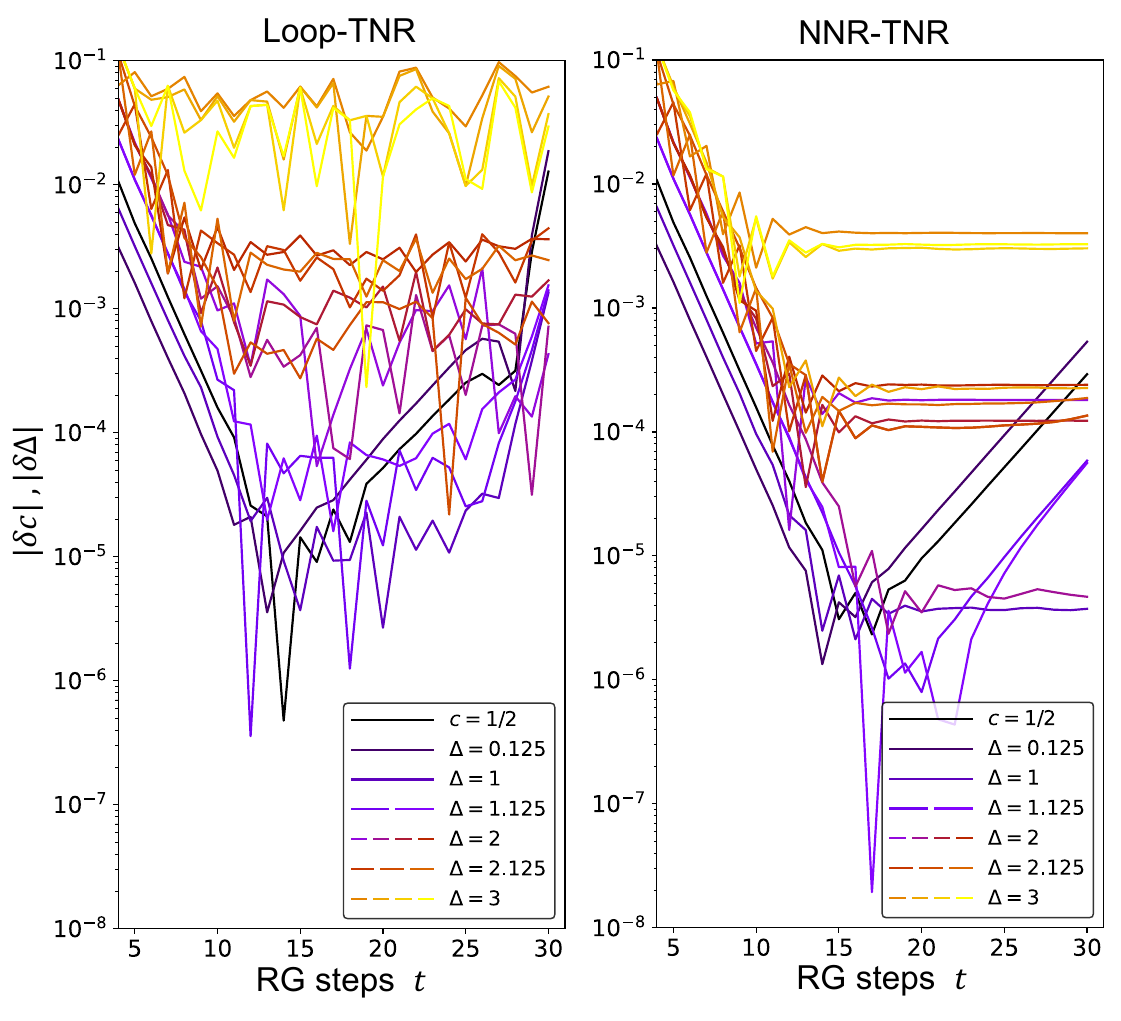}% Here is how to import EPS art
        \caption{\label{fig:scaling_error} The RG steps dependence of the absolute value of relative errors of central charge $c$ and scaling dimensions 
        $\Delta$ for Loop-TNR and NNR-TNR with bond dimension $\chi = 32$. In this figure, $\xi^{\text{init}}$ and $\rho$ were chosen to be $ 4\cdot 10^{-6}$ and $0.85$, respectively}
\end{figure}
In Fig.~\ref{fig:cft}, we show the results of central charge and the smallest 40 scaling dimensions $\Delta$ computed from diagonalizing transfer matrix \cite{PhysRevB.80.155131} 
for Loop-TNR and NNR-TNR. In the Loop-TNR case, the higher levels of scaling dimensions spectrum become unstable as a function of the RG steps and hardly distinguishable, for examples
$\Delta \geq 3$ for $\chi = 16$ and $\Delta \geq 4$ for $\chi = 32$, respectively. On the other hand, in the case of NNR-TNR, such instabilities of the aforementioned levels of scaling 
dimensions are not seen. NNR-TNR successfully extracts $\Delta= 3$, $\Delta=3.125$ scaling dimensions at $\chi=16$ and $\Delta=4$ and $4.125$ are recognizable at $\chi=32$.
Although there exist some spurious solutions such as the one around $\Delta\approx 3.3$ for $\chi=16$ and $\Delta\approx 3.7$ for $\chi=32$, we can eliminate these spurious solutions based on their stronger dependence on $\chi$ and $L$ than the true solutions.

In terms of the accuracies of CFT data, we compare the absolute value of relative errors of central charge $c$ and the scaling dimensions upto $\Delta =3$
at different system size in Fig.~\ref{fig:scaling_error}. One can see that the 
relative errors of all the quantities approach to the exact values and deviate afterwards, which is an expected behavior 
due to finite size and finite bond dimension effects \cite{ueda2023finitesize}.
For the CFT data corresponding to the lower indices of the fixed-point tensor, e.g. central charge and the scaling dimensions of primary 
fields, NNR-TNR produces slightly more accurate and stable result against RG steps than Loop-TNR. In addition, we have observed that NNR-TNR tends to significantly improve 
the accuracies of the higher levels of scaling dimensions compared to Loop-TNR, by an order of magnitude for $\Delta=3$, for example. It seems to suggest that NNR-TNR can more 
effectively eliminate short-range correlations than Loop-TNR, thereby reaching a more genuine critical fixed-point tensor.

In all cases above, 50 or less iterations were required for NNR loop optimization at each RG steps.

%\footnote[4]{Recently, it has been reported that tensor network algorithm has the "finite bond dimension effects" induced by finite correlation length from bond dimension.  The meaning of slopes in Fig are discussed in .}. 

% to obtain the results.
%This seems to suggests that NNR loop optimizations efficiently avoid the local minima originated from short-range correlation.

\section{Conclusions and discussions}
\label{disc}
In this paper, we proposed alternative loop optimization for circumventing the local minima related to short-range correlation that lives within tensor 
space. We have pointed out adding nuclear norm regularization to loop optimization resolves the short-ranged correlation problems and demonstrated that 
it further improves the RG flow of critical 2D Ising model and higher parts of scaling dimensions becomes clearly visible using NNR-TNR.

The major advantage of NNR-TNR over other loop optimization techniques lies in its efficiency in eliminating short-range correlations, allowing us to stably obtain
more accurate critical fixed-point tensors through a simple numerical procedure. For example, an iteration of the Loop-TNR consists of several steps, 
such as entanglement filtering and periodic MPS optimization, to generate correct fixed-point tensor. In contrast, 
the present paper shows that the NNR loop optimization alone is sufficient to remove the short-range correlations and further improves the scale-invariance 
of the critical fixed-point tensor. For pursuits of evaluating higher and precise scaling dimensions, there remains several options, such as increasing 
bond dimensions, length of columns for constructing transfer matrix, and using gauge symmetries of the model e.g. $Z_{2}$ \cite{PhysRevB.83.115125,PhysRevA.82.050301}, 
rotational and reflectional symmetries. For those future developments, the NNR loop optimization presented above would serve as the good starting point 
for even more accurate tensor network algorithms.

From technical point of view, it should be noted that hyperparameter tuning of NNR loop optimization remain an issue;  we chose the parameters $\xi$ and 
$\rho$ of NNR-TNR, such that the critical fixed-point tensor exhibits scale-invariance up to the largest possible system size. However, these parameters are often 
situation-dependent. How to automatically tune these parameters for general cases remain for future research.

Even more challenging and interesting subject for future works is to generalize the NNR loop optimization for higher dimensions. 
We believe this can be achieved by incorporating the idea of local bond matrix truncation, as introduced in Refs.\cite{PhysRevB.98.085155,PhysRevB.97.045111}, 
into NNR loop optimization. It would serve as an alternative method of removing redundant correlation efficiently in the 2D quantum or 3D classical tensor 
network algorithms such as PEPS \cite{PhysRevB.102.075147} and TRG \cite{PhysRevB.86.045139,PhysRevB.102.054432,kadoh2019renormalization}.
\\
\begin{acknowledgments}
We thank Qibin Zhao, Satoshi Morita, Kenji Harada, Xinliang Lyu, Atsushi Ueda, Hyun-Yong Lee and Katsuya Akamatsu for fruitful discussions. 
T. O. acknowledges the support from the Endowed Project for Quantum Software Research and Education, The University of Tokyo \cite{qsw}.
This work was supported by JST SPRING, Grant Number JP-MJSP2108, JST Grant Number JPMJPF2221, JSPS KAKENHI Grant Numbers 19H01809 and 23H01092, and partially by the joint project of Kyoto University and Toyota Motor Corporation, titled “Advanced Mathematical Science for Mobility Society". The computation in this work has been done using the facilities of the Supercomputer Center, the Institute for Solid State Physics, the University of Tokyo.
\end{acknowledgments}
\appendix
\section{Source code}\nonumber
\label{app1}
Our sample code for NNR-TNR is available on github.com/KenjiHomma/NNR-TNR. It can be used to reproduce all the results shown in Sect~\ref{results}.

\nocite{*}

\bibliography{apssamp}% Produces the bibliography via BibTeX.

%%%%%%%%%% Merge with supplemental materials %%%%%%%%%%
\clearpage

\pagebreak

\widetext
\begin{center}
\textbf{\large Supplemental Material: Nuclear norm regularized loop optimization for tensor network}
\end{center}
%%%%%%%%%% Merge with supplemental materials %%%%%%%%%%
%%%%%%%%%% Prefix a "S" to all equations, figures, tables and reset the counter %%%%%%%%%%
\setcounter{equation}{0}
\setcounter{figure}{0}
\setcounter{table}{0}
\setcounter{page}{1}
\makeatletter
\renewcommand{\theequation}{S\arabic{equation}}
\renewcommand{\thefigure}{S\arabic{figure}}
\renewcommand{\bibnumfmt}[1]{[S#1]}
\renewcommand{\citenumfont}[1]{S#1}
%%%%%%%%%% Prefix a "S" to all equations, figures, tables and reset the counter %%%%%%%%%%

\section*{Loop optimization based on entropy regularization}
In the main text, we choose to penalize the nuclear norm of $S_{[\alpha(n)]}$ to avoid the local minima. As we briefly discussed in the main text, there may be several other choices 
for the penalty term. Here we consider a term that has the same form as the Shanon entropy. Similar to the nuclear norm, it is a function of the singular values of the matricized 
tensors. Instead of their simple sum, we consider $-\sum_i \sigma_i \log \sigma_i$ with the normalization $\sum_i \sigma_i = 1$. This can be interpreted as the local mutual 
information between the extracted leg and the others (after neglecting the environment). We call the resulting regularization the entropy regularization (ER) hereafter. 
In ER, the cost function Eq.(\ref{eq.3}) is replaced by 
\begin{eqnarray}
    \label{eq.S1}
    \min_{\mathcal{S}} \frac{1}{2}||\ket{\psi(\mathcal{T})}- \ket{\psi(\mathcal{S})} ||^{2}_{F} + \lambda \sum_{n=1}^{N} E_{[\alpha(n)]}^{n} ,
\end{eqnarray}
where $E^n_{[\alpha(n)]}$ is the local mutual information mentioned above. In this setup, the hyperparameter $\lambda$ is set to be 
$\lambda =\tilde{\lambda} ||\ket{\psi(\mathcal{T})}- \ket{\psi(\mathcal{S})}||^{2}_{F}$ and $\tilde{\lambda}$ is a positive parameter for every RG steps.

Intuitively, penalizing the entropy over the diagonal legs in Fig.~\ref{fig:cost_function} should also induce the low-rankness and thereby lead to a reduction in redundant correlation, as does the NNR technique. 
The remaining question is which of the two approaches is numerically preferable for regularization technique in the context of TNR. 

The numerical setup for this demonstration, such as the initial condition of $S$ and normalization of $T$, is equivalent to the recipes discussed
 in main text. However, the updating scheme of $S$ and $\tilde{\lambda}$  is slightly different from the main text. 
 We elaborate the updating scheme at each RG step as follows: we perform 10 sweeps of nonlinear Conjugate Gradient method to optimize Eq.(\ref{eq.S1}) with $\tilde{\lambda} = 0.01$. Then, set $\tilde{\lambda} = 0$ and
  perform 50 sweeps of optimization using the periodic MPS method. Hereafter, we call the TNR based on the ER method as ER-TNR.

\begin{figure}[H]
    \centering
 
    \includegraphics[scale=0.32]{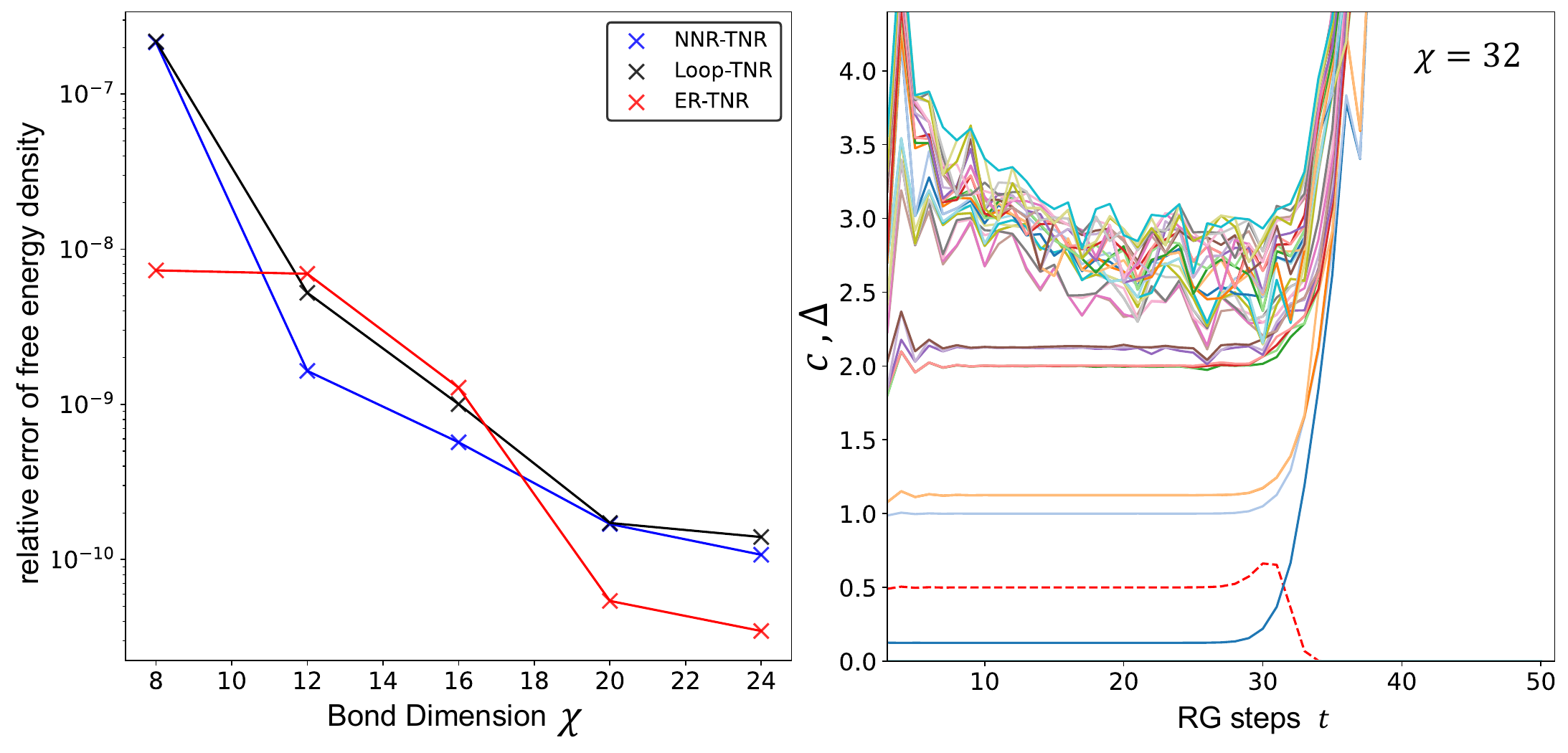}% Here is how to import EPS art
    \caption{\label{fig:ep}(Left) Relative error of free energy density of critical 2D classical Ising model as a function of bond dimension using 
    NNR-TNR, Loop-TNR and ER-TNR with $2^{30} \times 2^{30}$ spins. The blue solid line corresponds to NNR-TNR result with the parameters $\xi^{\text{init}}=1\cdot 10^{-5}$ 
    and $\rho=0.84$, chosen to be fixed over the full range of bond dimensions for simplicity. (Right) The conformal data extracted by ER-TNR at different RG steps with bond 
    dimension $\chi = 32$.
    }
    
\end{figure}

  Fig.~\ref{fig:ep}(Left) compares the relative error of free energy density of critical 2D classical Ising model with different bond dimensions among ER-TNR, NNR-TNR and Loop-TNR. It seems that ER-TNR gives comparable results with the other techniques to NNR-TNR or Loop-TNR for $\chi <20$, and is even more accurate for the higher bond dimension $\chi = 20, 24$. 
  However, the comparison between Fig.~\ref{fig:ep}(Right) and Fig.~\ref{fig:cft}(b)(d) shows that the stability of central charge and scaling dimension spectrum as a function of RG step seems to deteriorate around the 30th RG step,
   and the higher levels of scaling dimension spectrum could not be obtained for $\Delta>2$ in Fig.~\ref{fig:ep}(Right). In addition, we also find that the stability of singular value spectrum and the relative error of scaling dimension spectrum are rather inferior to the results shown in Fig.~\ref{fig:spectrum} and Fig.~\ref{fig:scaling_error}.

 From the perspectives of the numerical RG algorithm, the NNR seems to be more suitable than the ER, at least for loop 
optimization in TNR, although the proper “norm”  may depend on the given entanglement structure, e.g. the initial condition.

\end{document}